\documentclass[aps,pre,epsf,superscriptaddress,amsmath,amssymb,amsfonts,twocolumn,showpacs]{revtex4-1}

\usepackage{graphicx}
\usepackage{epsfig}
\usepackage{dcolumn}
\usepackage{bm}
\usepackage{braket}
\usepackage{amsmath}
\usepackage{color}

\begin{document}

\title{Collective excitations of dipolar gases based\\ on local tunneling in superlattices}

%\title{Collective excitations of dipolar gases based on\\ local tunneling in ultracold superlattices}

\author{Lushuai Cao}
\email{lushuai\_cao@hust.edu.cn}
\affiliation{Ministry of Education Key Laboratory of Fundamental Physical Quantities Measurements,
School of Physics, Huazhong University of Science and Technology, Wuhan 430074,People's Republic of China}
\author{Simeon I. Mistakidis}
\affiliation{Zentrum f\"{u}r Optische Quantentechnologien,
Universit\"{a}t Hamburg, Luruper Chaussee 149, 22761 Hamburg,
Germany}
\author{Xing Deng}
\affiliation{Ministry of Education Key Laboratory of Fundamental Physical Quantities Measurements,
School of Physics, Huazhong University of Science and Technology, Wuhan 430074,People's Republic of China}
\author{Peter Schmelcher}
\email{pschmelc@physnet.uni-hamburg.de}
\affiliation{Zentrum f\"{u}r Optische Quantentechnologien,
Universit\"{a}t Hamburg, Luruper Chaussee 149, 22761 Hamburg,
Germany} \affiliation{The Hamburg Centre for Ultrafast Imaging,
Universit\"{a}t Hamburg, Luruper Chaussee 149, 22761 Hamburg,
Germany}

%\ead{\mailto{$^*$ lushuai\_cao@hust.edu.cn},\\
 %    \mailto{$^{\S}$ pschmelc@physnet.uni-hamburg.de}}

\date{\today}

\begin{abstract}
The collective dynamics of a dipolar fermionic quantum gas confined
in a one-dimensional double-well superlattice is explored.
The fermionic gas resides in a paramagnetic-like ground state in the weak interaction regime, upon which
a new type of collective dynamics is found when applying a local perturbation.
This dynamics is composed of the local tunneling of fermions in separate supercells, and
is a pure quantum effect, with no classical counterpart. Due to the presence of the dipolar interactions
the local tunneling transports through the entire superlattice, giving rise to a collective dynamics.
A well-defined momentum-energy
dispersion relation is identified in the ab-initio simulations
demonstrating the phonon-like behavior. The phonon-like characteristic is also confirmed by an
analytical description of the dynamics within a semiclassical picture.

\end{abstract}

\maketitle
%\pacs{} \maketitle

\section{Introduction}

Collective excitations constitute a fundamental concept in condensed matter physics which is at the origin of
various phenomena in the field \cite{Anderson, Simon}. Remarkable examples of collective excitations
are phonons, magnons or plasmons. Among them, phonons describe the
collective dynamics of the atomic vibrations in the crystal lattice, and play a key role for different fundamental
effects in condensed matter physics, such as superconductivity \cite{Bardeen} or the thermal transport
in solid matter \cite{Lepri,Li}.

Generalizations of the concept of a phonon can be found in
ion traps \cite{Porras,Bissbort} or ultracold dipolar quantum gases \cite{Pupillo,Ortner}.
Phonons in ion traps refer to the collective dynamics of ions' motion around their equilibrium positions in e.g.
Paul traps. In dipolar quantum gases, it describes the coupling between the local vibrations of dipolar atoms
in a self-assembled chain or a lattice. Generally speaking, phonons in crystals, ion chains and dipolar lattices
all refer to the collective dynamics of vibrations, which have a direct analogue to the motion
of classical vibrators, and these phonons can be seen as a direct extension of the classical
dynamics of a vibrating chain to the quantum regime.
Here, we introduce a new type of collective dynamics in ultracold dipolar gases
in a one-dimensional superlattice. The key ingredient for this collective dynamics is a local tunneling,
which possesses no classical counterpart.

Our investigation is mainly based on ab-initio simulations,
besides a semiclassical analytical treatment. The simulations are performed by employing the numerically exact
Multi-Layer Multi-Configuration Time-Dependent Hartree method for identical particles and mixtures (ML-MCTDHX) \cite{MLX}, which
has been developed from MCTDH \cite{Meyer,Beck}, ML-MCTDH \cite{Wang,Manthe} and ML-MCTDHB \cite{Kronke,Cao},
and has a close relation to MCTDHB(F) \cite{Alon,Alon1,Axel}. The ab-initio simulations of
the corresponding ultracold quantum gases take into account all correlations, and can unravel new effects beyond the
predictions of mean-field theory and for lattice systems beyond the single-band Bose-Hubbard model.
Representative examples along this line are the loss of coherence and the decay of contrast of different
types of solitons \cite{Streltsov,Kronke1}, higher band effects on the stationary \cite{Alon2}
or dynamical properties \cite{Zollner,Sakmann,Cao1,Mistakidis,Mistakidis1,Mistakidis2} in optical lattices.
To investigate the collective dynamics in the double-well superlattice, we employ ML-MCTDHX which allows for a full description of the
dynamics, and proves the robustness of the collective dynamics against higher order correlations and higher
band effects.

This work is organized as follows: In section II, we present
an introduction to the detailed setup (Sec. II.A), the initial state preparation (Sec. II.B), the local effect
of the perturbation that drives the system out of equilibrium (Sec. II.C), and the global collective dynamics
induced by the local perturbation (Sec. II.D). We also supply a semiclassical analytical
description of the collective dynamics (Sec. II.E). The discussion of our results and the conclusions are provided in section III.

\section{Collective excitations based on local correlation-induced tunneling}

\subsection{Setup}

We consider a dipolar superlattice quantum gas (DSG) composed of $N$ spin-polarized fermions confined
in a one-dimensional double-well superlattice of $N$ supercells, $i.e.$ a unit filling per supercell.
All the fermions interact with each other by dipolar interactions. The Hamiltonian read as follows
\begin{equation}\label{ham}
 H=\sum_{i=1}^{N}(\frac{-\hbar^2}{2M}\partial^2_{x_i}+V_{sl}(x_i))+\sum_{i<j=1}^N\frac{D}{\mid x_i-x_j\mid^3+\delta}.
 %+\sum_{i=1}^{N}V_{tr}(x_i,t).
\end{equation}
The first term refers to the single-particle Hamiltonian, where $V_{sl}$ models the double-well
superlattice with $V_{sl}(x)=V_0(sin^2(kx)+2cos^2(2kx))$. This superlattice can be formed by two pairs of
counter-propagating laser beams of wave vectors $k$ and $2k$, and the strength of the lattice $V_0$ can be tuned
by the amplitude of the laser beams. We consider a finite-length lattice of $N$ supercells, and hard-wall boundaries are
applied at positions $x=\pm N\pi/(2k)$, to allow only $N$ supercells in our simulation. The second term in the Hamiltonian
models the dipolar interaction between the fermions. In this work we consider the situation that all the dipoles are
polarized along the same direction, perpendicular to the relative distance between the fermions, and $D$
denotes the strength of the interaction. To avoid in simulations the divergence of the interaction
at $x_i=x_j$, an offset $\delta$ is added to the denominator. The offset $\delta$ takes a rather small value, 
which is about eight times smaller than the spacing of the discrete grid points chosen 
within our simulations. More specifically, in the present work we focus on the situation where all fermions reside 
in different wells for which case the distances where a significant overlap exists are much larger than $\delta$. Then, the offset 
is negligible. 

\begin{figure}
\includegraphics[width=8cm]{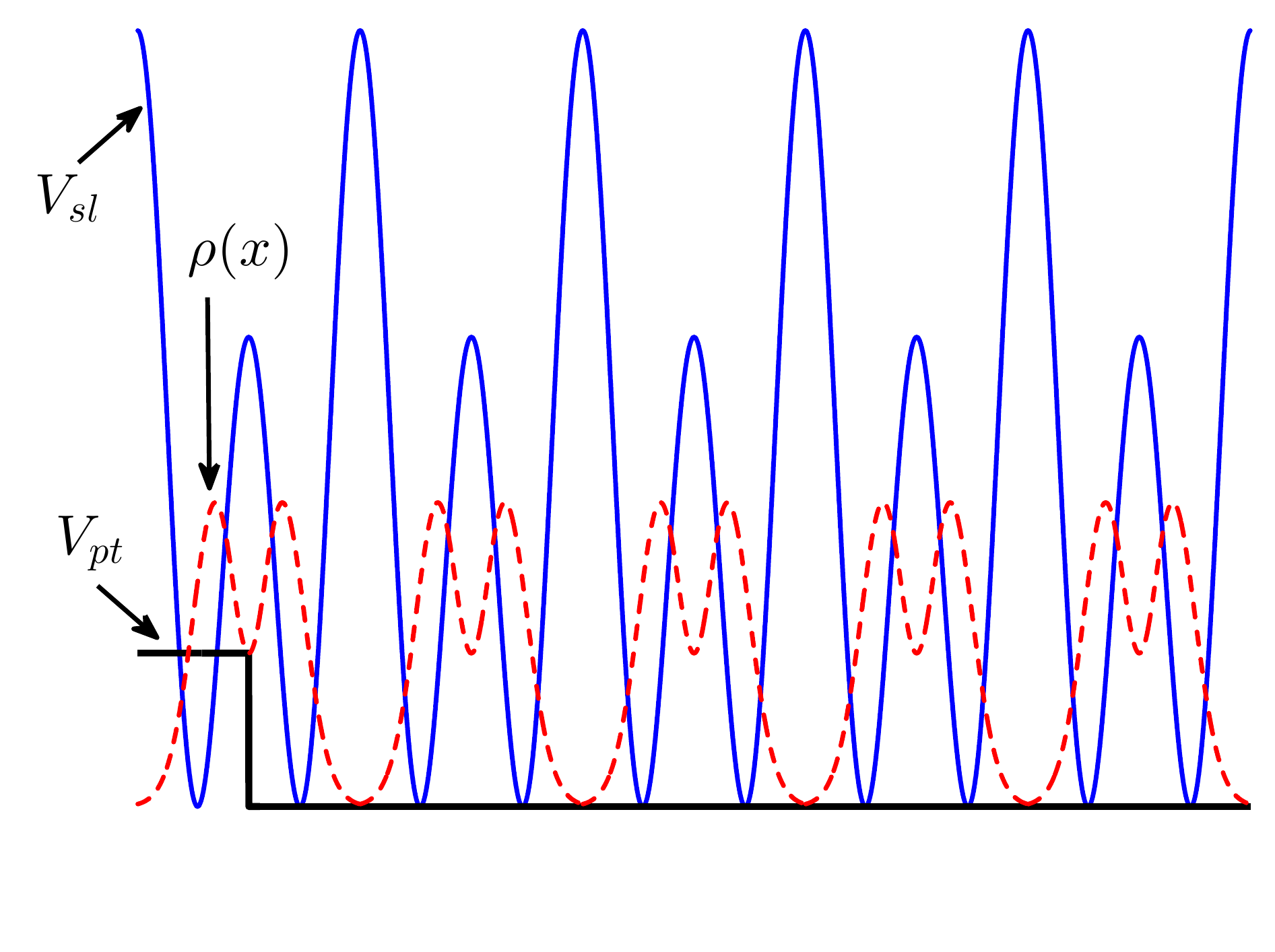}
\caption{Sketch of the dipolar fermionic gas in a double-well superlattice of five cells. The blue and black lines show the double-well
superlattice and the local perturbation modeled as a step function applied only to the
most left site of the lattice, respectively. Five fermions (red dashed gaussians) are loaded to the superlattice, each of which is
localized in a different supercell and occupies the two sites in the cell for the initial state.} \label{fig1}
\end{figure}

In order to investigate the collective excitations, the DSG is firstly relaxed to the ground state of $H$,
and at $t=0$ a local perturbation is applied to a single supercell of the lattice, $e.g.$ the outer most left cell is taken
out of equilibrium. This perturbation is intended to induce a local dynamics in the left
cell, and is applied only for a short time period, to avoid affecting the global dynamics on a long time scale.
We model the local and temporal perturbation by $V_{pt}(x,t)=V_1~\theta(x+(N-1)\pi/(2k))~\theta(t-\tau)$, with the Heaviside step function
$\theta(x)$. The perturbation is then modeled as a local step function applied to
the outer most left supercell and lasts only for the temporal interval $[0,\tau]$. The double-well superlattice and the local perturbation
are sketched in figure \ref{fig1} for five fermions in a five-cell superlattice.
In the simulation, we render the Hamiltonian dimensionless by setting $\hbar=M=k=1$, which is equivalent to rescaling the energy,
space and time in the units of $E_R=\hbar^2k^2/2M$, $k^{-1}$ and $\hbar/E_R$, respectively.

The setup discussed above can be realized in ultracold atom experiments. Moreover, dipolar quantum gases have become
a hot topic in the field of ultracold atoms and molecules \cite{Lahaye,Baranov}.
Their rich phase properties \cite{Yi,Capogrosso,Hauke,Kadau,Barbut}
and perspectives in, for instance,
quantum simulations \cite{Micheli,Gorshkov,Kaden} have inspired extensive studies on
dipolar quantum gases. Experiments can nowadays
prepare dipolarly interacting particles in lattices, due to the rapid development of cooling
atoms \cite{Zhou,Olmos,Baier} with large magnetic dipole moments and polar
molecules in optical lattices \cite{Ni,Deiglmayr,Yan,Guo,Frisch}.
Specifically, the double-well superlattice has been
realized in experiments, and has become a widely used testbed for various phenomena, such as correlated atomic tunneling
\cite{Folling}, generation of entanglement of ultracold atoms \cite{Dai} and the topological
Thouless quantum pump \cite{Lohse}. The setup discussed and analyzed here is therefore well within experimental reach.

\subsection{Paramagnetic-like initial state}\label{initial}
The dynamics investigated in the present work strongly depends on the choice of the initial state, which is prepared as
the ground state of $H$ in a particular parameter regime.
It has been shown \cite{Yin} that the DSG system can be mapped to an effective Ising spin chain model under the so-called pseudo-spin mapping. 
Then, within the Ising spin picture the ground state of the system undergoes a transition from a
paramagnetic-like state to a single-kink state  for increasing dipolar interaction. 
The paramagnetic-like state refers to the pseudo-spins polarized in the same direction due to an effective
magnetic field, whereas the single-kink state is composed of two effective ferromagnetic domains aligning in 
opposite directions. 
In the present work, we focus on the dynamics in the weak interaction regime, $i.e.$ the DSG system initially
resides in the paramagnetic-like state.

To comprehend and analyze qualitatively the initial particle configuration (being characterized by the many-body state $\ket{\Psi}$) we shall employ the notion of
reduced densities. The one-body reduced density matrix $\rho_1(x,x')=\braket{x'|\hat\rho_1|x}$,
is obtained by tracing out all fermions but one in the one-body density operator $\hat\rho_1\equiv tr_{2,...,N}\ket{\Psi} \bra{\Psi}$ of
the $N$-body system, while the two body density $\rho_2(x_1,x_2)=\braket{x_1,x_2|\hat\rho_2|x_1,x_2}$ can be obtained by a partial trace over all but two
fermions of the two-body density operator $\hat\rho_2\equiv tr_{3,...,N}\ket{\Psi} \bra{\Psi}$.
Subsequently, the initial state can be characterized by the two-body and one-body correlations in the superlattice, as shown in figure 2.
Figure 2($a$) presents the two-body correlation of five fermions in a five-cell superlattice. The vanishing occupation along the
diagonal direction in the two-body correlation illustrates that no two fermions (or more) occupy the same supercell, and each
supercell hosts only one fermion, which is a Mott-like configuration. Being localized in a separate supercell,
the fermions can occupy the left and right sites of the cell simultaneously, giving rise to particle number fluctuations in these sites and in particular to a non-vanishing off-diagonal
one-body correlation, as shown in figure 2($b$). This non-vanishing one-body correlation plays a key role in the collective
dynamics investigated in the present work.

To a good approximation, the paramagnetic-like ground state (see also Appendix B) can be expressed as
\begin{equation}\label{ground}
|G\rangle=\sqrt{2^{-N}}\prod_{i=1}^N(|L\rangle_i+|R\rangle_i),
\end{equation}
where $|L\rangle_i$ and $|R\rangle_i$ denote the lowest-band Wannier states in the left and right site of the $i$-th supercell,
respectively.
\begin{figure}
\includegraphics[width=7cm]{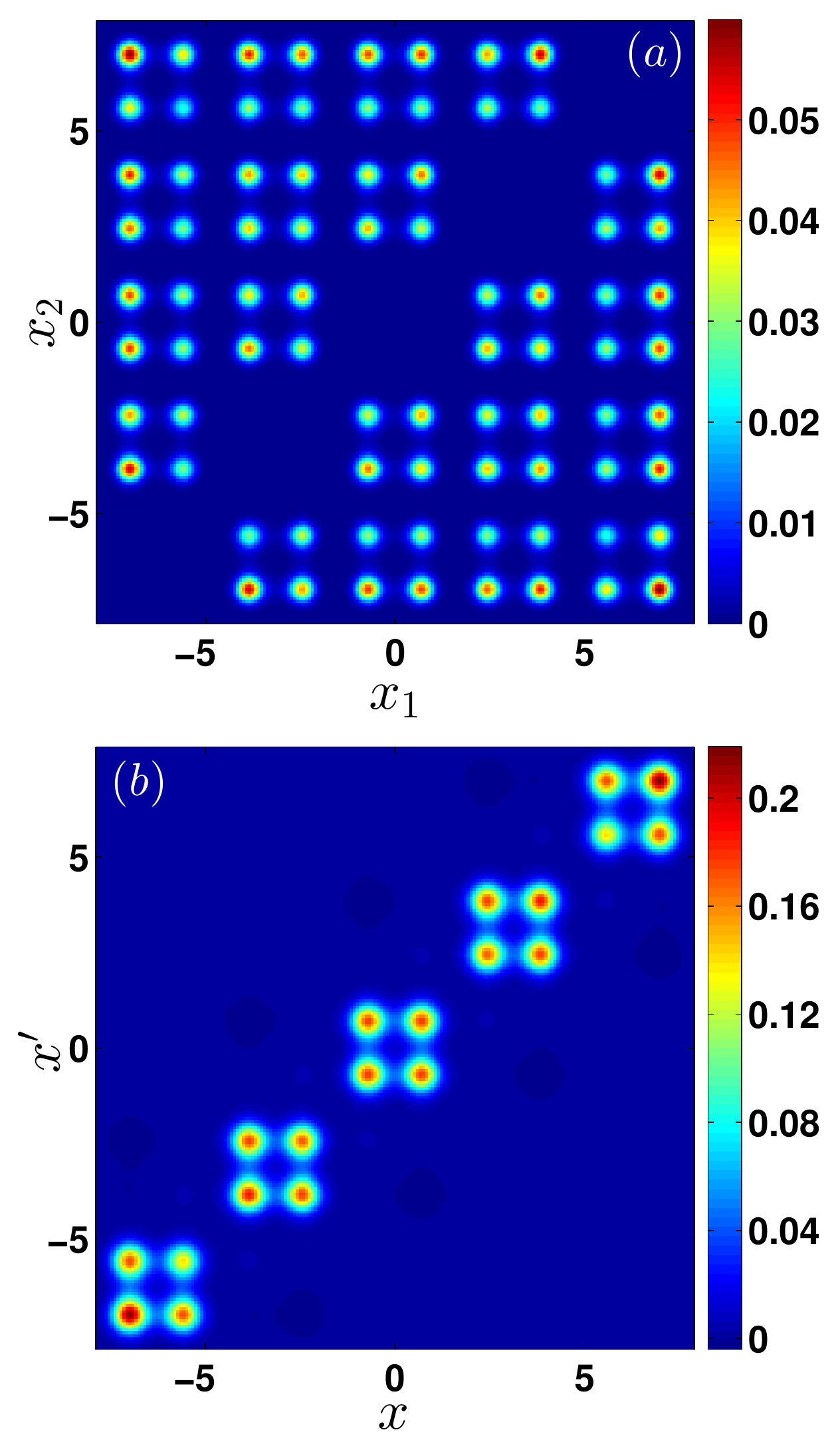}
\caption{($a$) The two-body correlations and ($b$) one-body correlations of the initial state of five fermions
in a five-cell superlattice. The two-body correlation illustrates that any two fermions cannot occupy the same
supercell, $i.e.$ each supercell hosts a single fermion. The off-diagonal of the one-body
correlations indicates the delocalization of the fermion between the left and right sites of the supercell.
The parameters used here are $(V_0,D)=(10,0.3)$, which correspond to a weakly interacting dipolar gas in a deep superlattice.
} \label{fig2}
\end{figure}

\subsection{Correlation-induced tunneling in a single supercell}

To drive the system out of equilibrium from the initial state $|G\rangle$, we apply a local perturbation $V_{pt}$
to the most left supercell. The perturbation is intended to induce a local tunneling of the fermion in
this supercell and is modeled by a step function applied to the left site of the supercell. In this subsection we
describe the local dynamics in this supercell under the perturbation $V_{pt}$.
The step function introduces an energy offset between the left and right sites of the cell. Normally,
the energy offset inhibits the tunneling between the two sites, of which the amplitude is reduced, by increasing the
amplitude of the offset. When a particle is initially prepared in a superposition state involving the two sites equally,
however, the offset can enhance the tunneling of the particle in a narrow parameter window of the offset strength.
The tunneling amplitude becomes maximal when the strength of the perturbation matches that of
the hopping between the two sites.

\begin{figure}
\includegraphics[width=8cm]{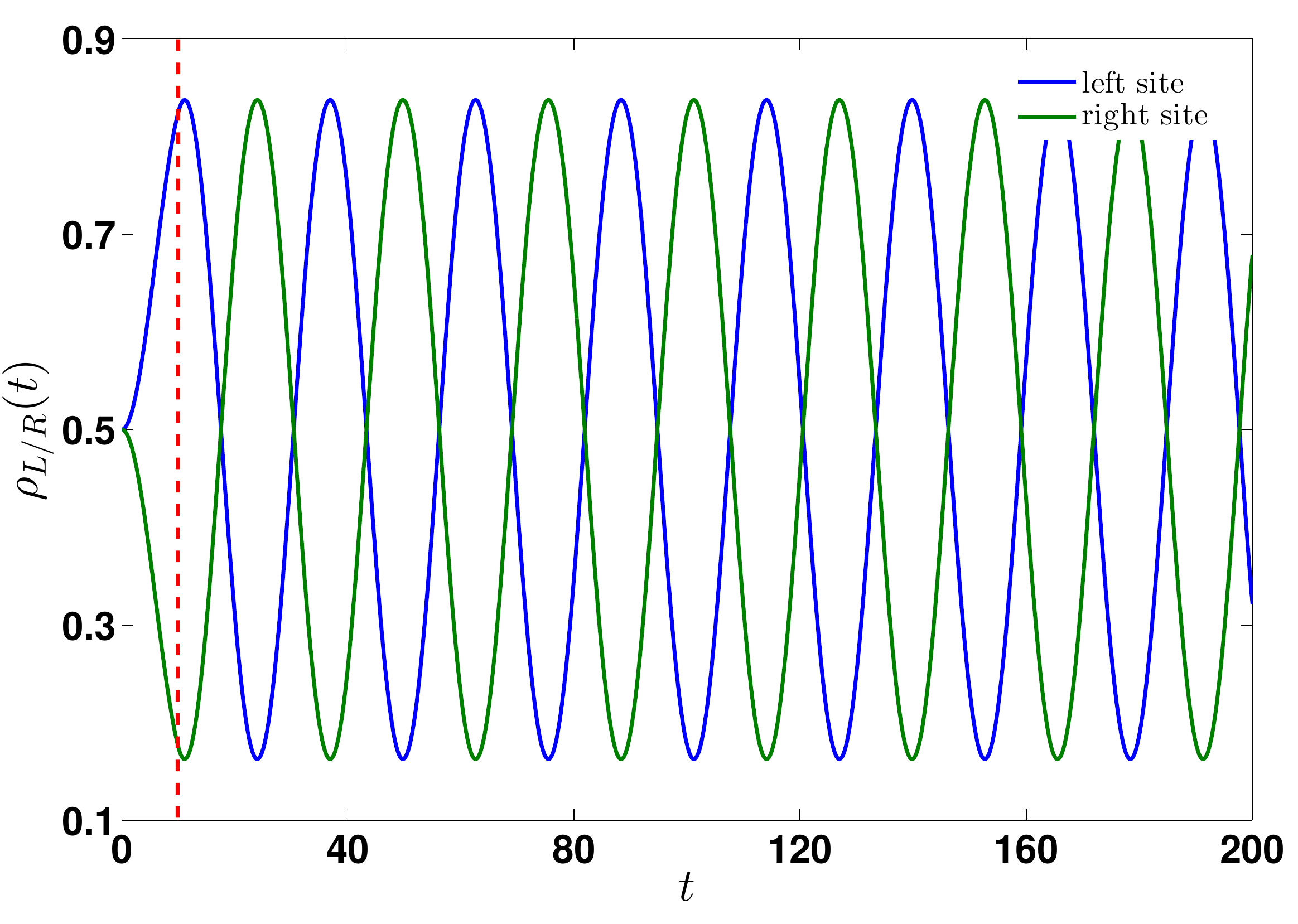}
\caption{Density oscillation of a single particle in a double well with a local perturbation applied to the left well.
The local perturbation is applied for a short time period, and the red dashed line marks the time when it is turned off.
The double well is taken from a single unit cell of the superlattice with $V_0=10$, and the height of the perturbation
potential is $h_1=0.1$.} \label{fig3}
\end{figure}

The explanation of such an unusual tunneling is as follows: In the normal case, the energy offset breaks the resonance between
the two sites in terms of the potential energy, and thus it inhibits the tunneling between the two sites.
When the initial state is chosen as a superposition state of the particle occupying the two sites equally,
a finite kinetic energy is stored in the system. The finite kinetic energy can then compensate the resonance breaking of
the potential energy and promote the tunneling. The maximum compensation is reached when the energy
offset matches the initial kinetic energy, which can be realized when the strength of the offset equals the hopping strength. 
In the double well system, the kinetic energy coincides with the one-body correlation between the two sites, up to a factor
determined by the hopping strength, and we term this unusual tunneling as correlation induced tunneling (CIT) \cite{Cao2},
to indicate the connection between the kinetic energy and the one-body spatial correlation. 
Moreover, the CIT can also be viewed as a Rabi oscillation between the two states
$(|L\rangle\pm|R\rangle)/\sqrt{2}$, where the tilt couples these two states and determines the corresponding Rabi frequency. 
In figure \ref{fig3} we illustrate the CIT of a single particle confined in a double well potential with a temporal
energy offset.
To proceed we calculate the population of each well, e.g.
for the right well $\rho_R(t)=\int_{0}^{\pi}dx \rho_1(x,t)$ (with $\rho_1$ being the one-body density).
As shown in the figure, initially the particle is occupying both sites with equal probability,
and after the perturbation (applied at $t=0$), the probability oscillates from the right to the left well,
indicating a tunneling between the two wells.
When the perturbation is turned off (the turn-off time is marked by the dashed red line in figure 3), we observe that
the tunneling persists.
Turning to the whole superlattice, it can be expected that the CIT also takes place in the left supercell when the same perturbation
is applied to a double-well supercell.

\subsection{Collective dynamics of local CIT}
Having introduced the initial state and the local dynamics of the CIT, let us proceed to the global dynamics of the entire
DSG system being subjected to a local perturbation. Our main finding can be summarized as follows: Once the local perturbation induces the CIT in
a single supercell, $e.g.$ the most left one as considered here, the dipolar interaction can transport the local CIT to other
cells. In this manner, all the fermions, while remaining well localized in their separate supercells, perform local CIT between
the two sites of their supercells,
giving rise to a collective dynamics of local CIT in the DSG system. Moreover, the collective dynamics resembles
phonon-like excitations, with a well defined momentum-energy dispersion relation. In following, we shall
demonstrate the collective dynamics of local CIT employing ab-initio simulations.
\begin{figure}
\includegraphics[width=8cm]{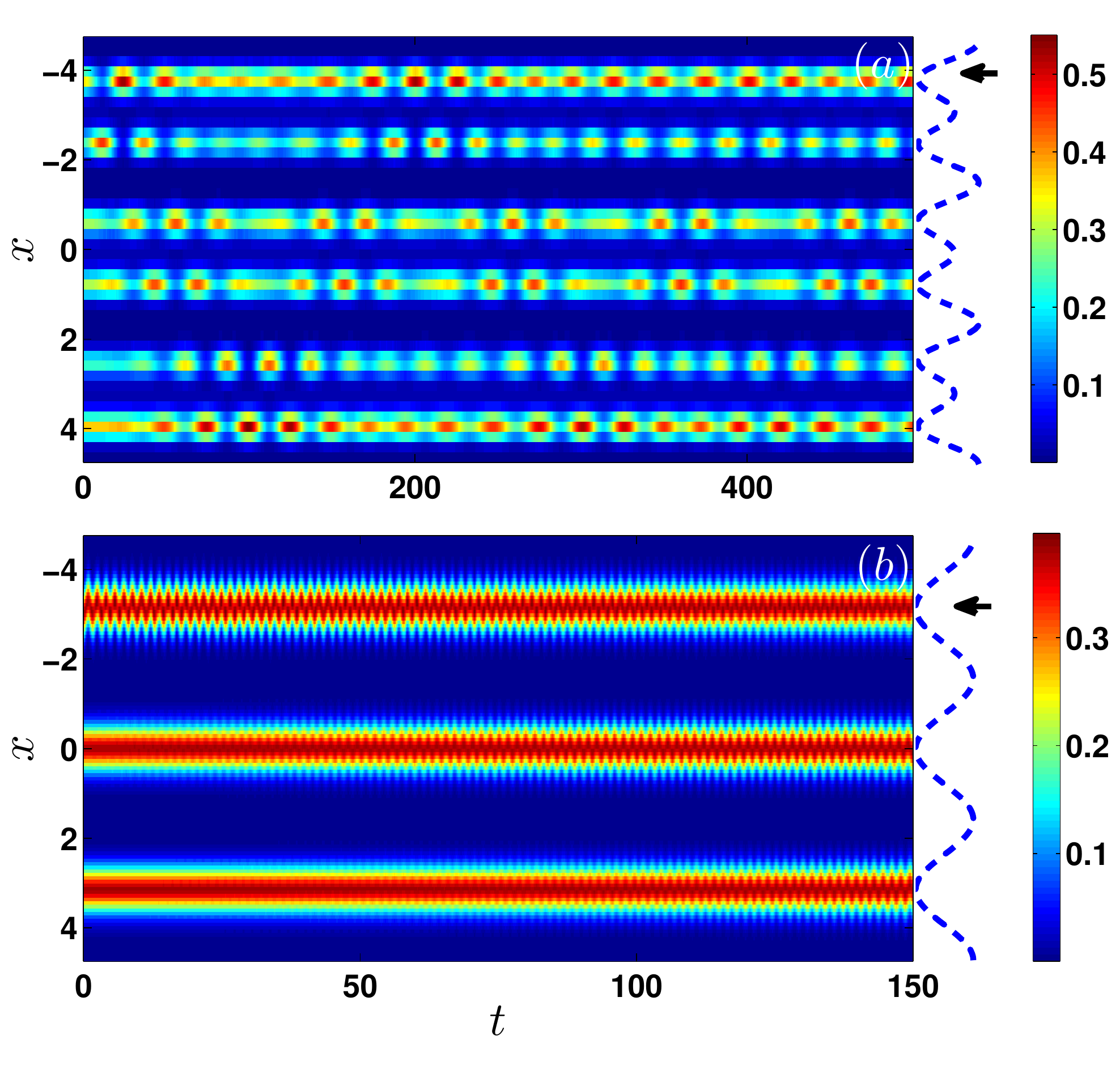}
\caption{($a$) The density evolution $\rho_1(x,t)$ of a three-fermion system in a three-cell superlattice, and ($b$) in a plain triple well,
under a corresponding local perturbation. The dashed lines attached to the right of the figures illustrate the corresponding trapping potentials. The arrows
mark the position where the local perturbation is applied for both cases. The parameters used here and in figure 5 are $(V_0,D,V_1)=(10,0.3,0.1)$.} \label{fig4}
\end{figure}

Firstly, we simulate the collective dynamics of $N=3$ fermions confined in a three-cell superlattice, $i.e.$ a 3F3C (3 fermions in 3 cells) system,
and compare it with the phonon of three dipolar-interacting fermions in a plain triple well. In figure 4($a$)
we show the one-body density oscillation $\rho_1(x,t)$ of the 3F3C system under the perturbation $V_{pt}$.
We observe that CIT takes place in all the three supercells, with no inter-cell tunneling between neighboring supercells.
This collective dynamics of CITs is different from the dipolar phonon as well as the ion phonon, which refer to the collective
dynamics of local classical vibrations of dipolar atoms or ions confined in a lattice respectively. 
In figure 4($b$) we also present the one-body density of the dipolar phonon of three fermions in a plain triple well.
In the dipolar phonon case, a local tilt induces a dipole oscillation of the fermion in the left well,
and the dipolar interaction transports the local density oscillation to fermions in remote wells, giving rise to
the collective phonon dynamics. Firstly, a similarity can be drawn between the collective CIT and the dipolar phonon,
where both cases are composed of local dynamics coupled by the dipolar interaction. On the other hand, the distinction of the two collective
dynamics is also obvious: The dipolar phonon (as well as the ion phonon) is composed of local oscillations of particles
and can be seen as a direct extension of the classical phonons to the quantum regime.
Meanwhile, the collective dynamics of CIT has no counterpart in the classical world and is a pure quantum effect.
\begin{figure}
\includegraphics[width=8cm]{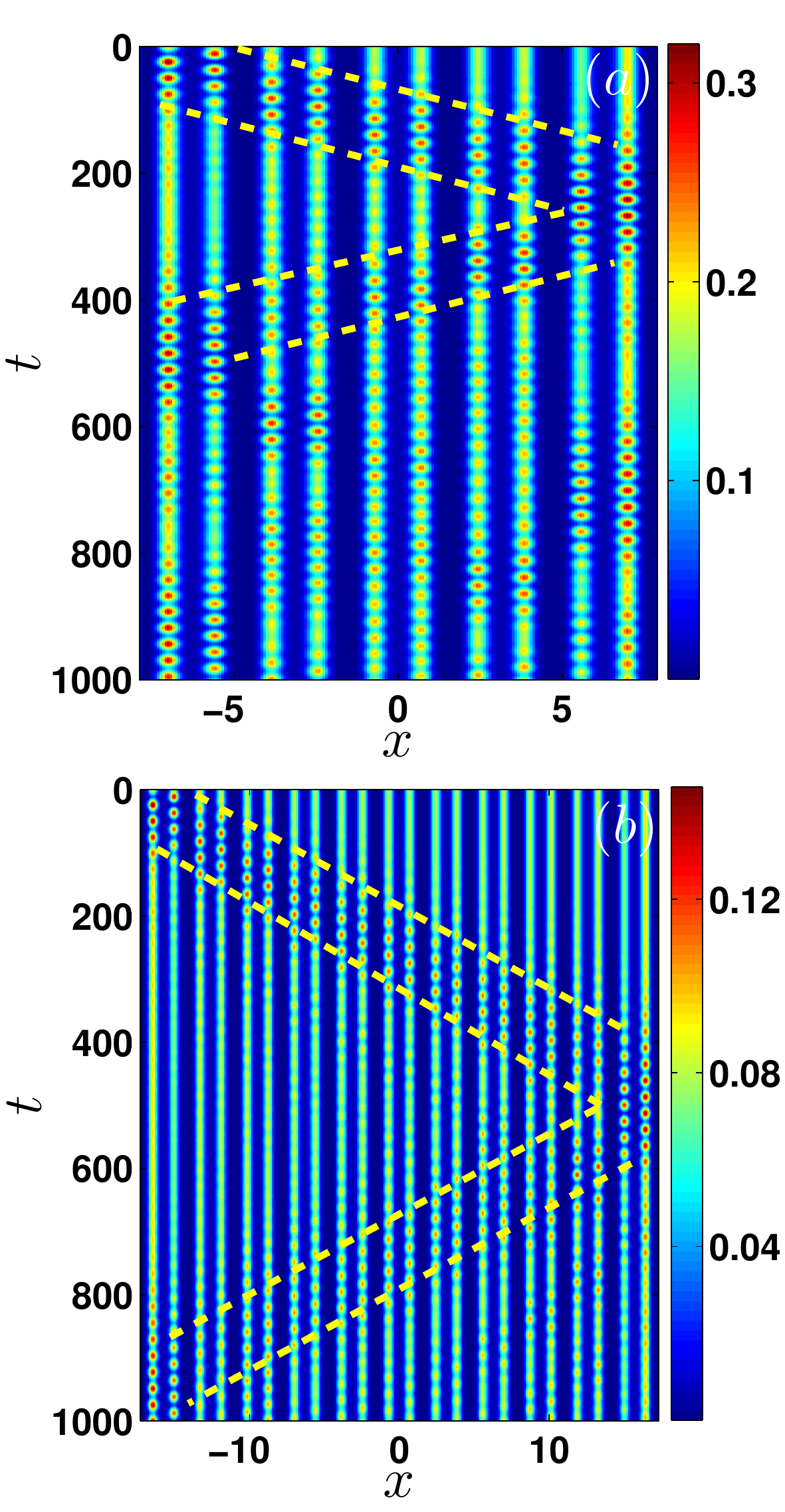}
\caption{The density evolution $\rho_1(x,t)$ of ($a$) a five-fermion and ($b$) eleven-fermion system of unit-filling, under the local perturbation.
The dashed yellow lines in both figures illustrate the finite transport velocity of the local CIT through the superlattice. 
The dashed lines (see the corresponding slopes) also suggest an equal transport velocity of the CIT in both systems,
implying that the transport velocity is independent of the system size. 
} \label{fig5}
\end{figure}

To demonstrate the generality of such collective dynamics with respect to the size of the superlattice we show that
the same behavior is evident in 5F5C (5 fermions in 5 cells) and 11F11C (11 fermions in 11 cells) systems, as shown in figures 5($a$) and 5($b$), respectively. In both figures we observe that
the collective dynamics of local CIT indeed takes place in bigger systems, indicating that it is not restricted
to a particular size. Moreover, in the longer lattices, we observe more clearly how the local CIT transport
through the whole system: They are not simultaneously excited along the lattice once the perturbation is applied,
but the CIT are transported with a finite velocity from the left supercell to remote ones. The transport of local CIT with a finite velocity
is illustrated in both figures with the yellow dashed lines, where one can even observe the reflection at the edges of the lattice.
In this way, the collective dynamics of local CIT in the DSG systems also serve as a test bed for the light-cone like behavior
of two-body correlations.

It is known that all phonon-like collective excitations share a common property of well defined momentum-energy dispersion
relation, where the collective dynamics can be decomposed into a set of momentum modes and each mode has a well defined
energy, $i.e.$ characteristic frequency. It is interesting to investigate whether the collective dynamics of DSG systems is also
associated with a dispersion relation. For this purpose, we calculate the density difference between the left and right
sites of each supercell $\delta\rho(i,t)$ ($i \in [1,N]$), and further define a set of $k$-modes as
\begin{equation}
 \delta\tilde\rho(k,t)=\sum_{n=1}^N \sin(\frac{nk\pi}{N+1})\delta\rho(n,t), k\in[1,N]
\end{equation}
To verify the corresponding dispersion relation we then calculate the spectra of $\delta\rho(i,t)$
and $\delta\tilde\rho(k,t)$. We show the spectra of
$\delta\rho(1,t)$ for 5F5C and 11F11C in figures 6($a_1$) and 6($b_1$), respectively, and the corresponding
spectra of $\delta\tilde\rho(k,t)$ in figures 6($a_2$) and 6($b_2$). These figures demonstrate that, firstly the
spectra of $\delta\rho(i,t)$ show $N$ main peaks for the $N$-fermion system, each of which corresponds to one $k$-mode,
indicating that the collective dynamics can be indeed decomposed into $N$ $k$-modes. More importantly, each $k$-mode is associated
with a dominant frequency peak, as shown in the spectra of $\delta\tilde\rho(k,t)$, and this directly verifies a
well-defined momentum-energy dispersion relation in the collective dynamics of the CIT. Further, we also observe
some weakly pronounced peaks lying near zero in the spectra, which are close to the values of the frequency difference between the corresponding peaks.
These peaks are attributed to a weak nonlinear effect similar to phonon-phonon interactions.
\begin{figure*}
 \centering
\includegraphics[width=14cm]{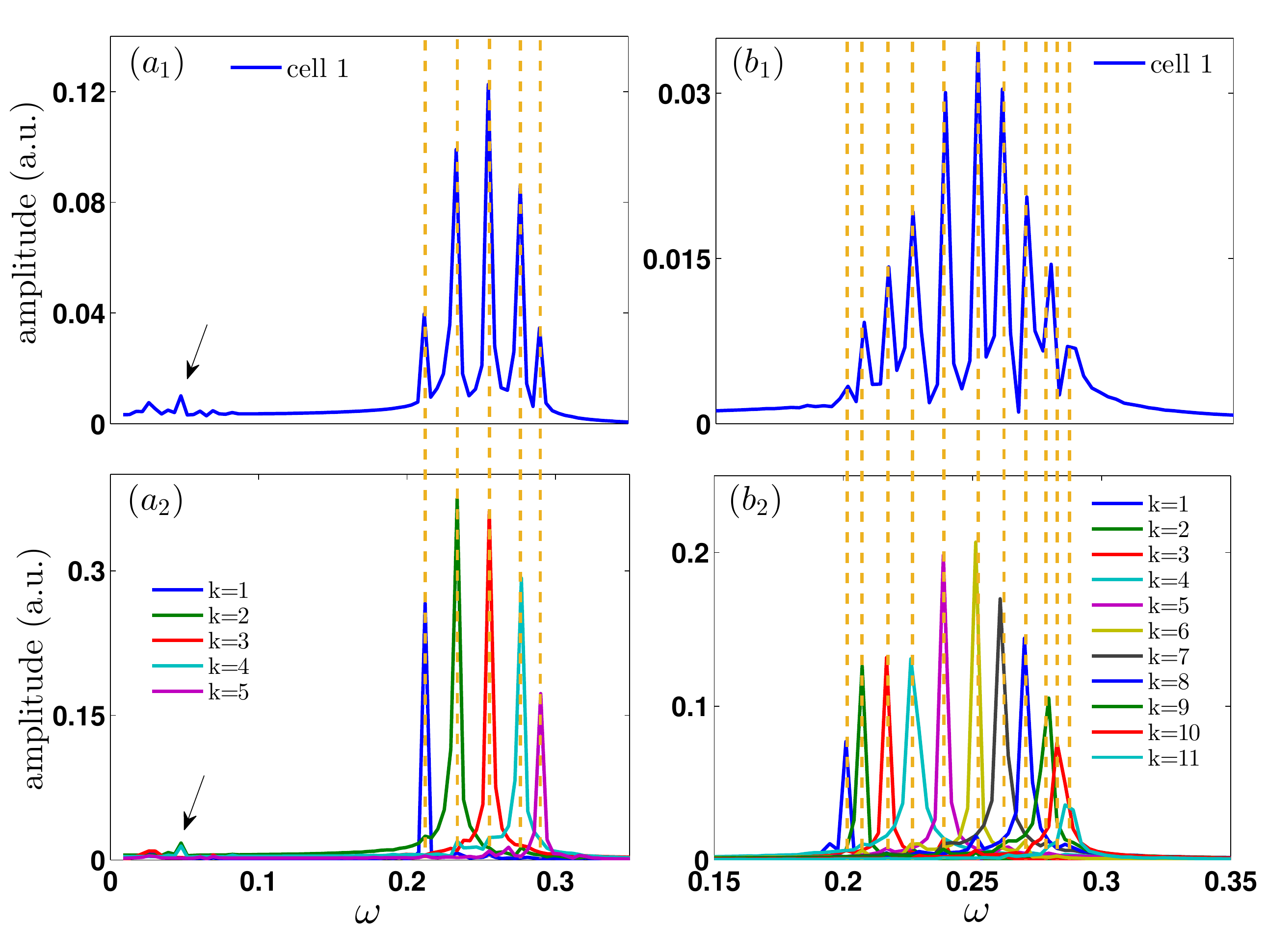}
\caption{The frequency spectra for a five-fermion (left column) and eleven-fermion (right column)
system of unit-filling. The upper and bottom rows correspond to the spectra of $\delta\rho(1,t)$ and
the $k$ modes $\delta\tilde\rho(k,\omega)$, respectively. The dashed vertical lines
demonstrate a one-to-one correspondence between the peaks in $\delta\rho(1,t)$ to the peak of a particular $k$-mode peak.
The arrows in figures ($a_1$) and ($a_2$) mark the tiny peaks in the low-frequency regimes, which are understood as a nonlinear effect of frequency subtraction.
As a result of the low resolution of the spectra with respect to
the dense packing of the peaks in the 11F11C case, some peaks are not well presented
in figure ($b_1$).} \label{fig6}
\end{figure*}

\subsection{Semiclassical description of the collective CIT dynamics}

In this section, we supply a semiclassical description of the collective CIT excitation, in terms of $\langle\delta\rho(i,t)\rangle$.
The starting point is the second-order time derivative equation
\begin{equation}\label{eqrho}
 -\partial^2_t\langle\delta\hat\rho(i,t)\rangle=\langle[[\delta\hat\rho(i),\hat H],\hat H]\rangle,
\end{equation}
which is derived simply by applying $i\partial_t\langle\delta\hat\rho(i,t)\rangle=\langle[\delta\hat\rho(i),\hat H]\rangle$ twice,
while the notation $\langle...\rangle$ denotes the expectation value $\langle\Psi(t)|...|\Psi(t)\rangle$. Then the major
task of solving equation (4) is to find proper expressions of the Hamiltonian $\hat H$ (for more details see Appendix B and
in particular equation (B1)) and to solve for $|\Psi(t)\rangle$.

We adopt the lowest-band Hubbard model for $\hat H$ and apply degenerate perturbation theory to solve for $|\Psi(t)\rangle$ and
derive a set of closed equations for $\langle\delta\hat\rho(i,t)\rangle$: a detailed derivation is given in Appendix B.
The final form of the equations that $\langle\delta\hat\rho(i,t)\rangle$ obeys, reads
\begin{equation}\label{rho}
\begin{split}
 -\partial^2_t\langle\delta\hat\rho(i,t)\rangle=4J^2\langle\delta\hat\rho(i,t)\rangle+4JV\big(\langle\delta\hat\rho(i-1,t)\rangle\\
 +\langle\delta\hat\rho(i+1,t)\rangle\big),
 \end{split}
\end{equation}
where $J$ and $V$ refer to the intra-cell hopping and the dipolar interaction strength, respectively.
Equation (5) is the semiclassical version of equation (4), and it is clear that equation (5) resembles that of
classical vibrating chains, where $\langle\delta\rho(i,t)\rangle$ plays the role of the local displacement of the $i$-th vibrator.
The general solutions of equation (5) correspond to a set of eigenmodes, and a particular solution is given by the superposition
of these eigenmodes (equation corrected) 
\begin{equation}\label{kmode}
\langle\delta\rho(i,t)\rangle=\sum_{k=1}^N \big(C_ksin(\omega_k t)+D_kcos(\omega_k t)\big)sin(\frac{ki\pi}{N+1}),
\end{equation}
where $\omega_k=\sqrt{4J[J+2Vcos(k\pi/(N+1))]}$, and $C_k$, $D_k$ are determined by the initial state.
The semiclassical equations (5) and their eigenmode solutions (see equation (6)) directly illustrate the phonon-like behavior of
the collective dynamics of the local CIT in the DSG system.

\section{Discussion and conclusions}

In this work we demonstrate a new type of collective excitations in dipolar quantum gases confined in the double-well
superlattice with a unit filling factor.
The collective excitations manifest themselves as the coupling and transport of local CIT within each supercell.
The local CIT are a pure quantum effect and have no classical counterpart, which endows
the dynamics composed of these collective excitations with a pure quantum nature, instead of being a quantum correction to
any classical dynamics.

These collective excitations can also be generalized from the double-well superlattice to more
complicate superlattices, where new properties of the collective dynamics can be engineered. For instance, the CIT in
a double well possesses a single characteristic frequency, and in the spectrum of the collective dynamics a single band arises from
this characteristic frequency. When the supercell is expanded to multiple wells, the corresponding characteristic frequencies of the local
CIT will also increase, and each of these frequencies seeds a band in the spectrum of the collective dynamics of the
local CIT, resulting in a multi-band spectrum.
The tunability of the band structure by the supercell properties indicate a high flexibility in designing
and engineering new properties of such collective excitations. Meanwhile, in the relatively strong interaction regime,
one can expect more pronounced nonlinear effects, such as the scattering of the collective excitations, which, however,
is beyond the scope of the current work.

It is in place to discuss the realizability and robustness of the collective excitations under realistic conditions.
Firstly, these excitations are not restricted to fermionic dipolar gases, but can also be realized with bosonic dipolar
gases, as the particles are localized in separate cells and the particle statistics plays almost no role here. For realistic
implementations, the collective excitations may be blurred by effects due to finite temperature, an additional external
potential and the imperfectness of the filling factor. To observe collective excitations, it is required to cool
the particles to the lowest band of the lattice. In previous experiments on double well superlattices, this condition has
been fulfilled for contact interacting atoms, and with the fast progress in cooling dipolar lattice gases we expect this
condition will become also feasible for our setup. In experiments, the confinement of lattice gases to a finite spatial domain
is realized by an external harmonic trap, which will also introduce some constraints on the realization.
In the bottom of the harmonic trap, it is possible to prepare a paramagnetic-like state, while at the edge deviations from
the perfect paramagnetic-like configuration can arise. It has been shown that this edge effect will not change the global
paramagnetic-like configuration \cite{Yin}, and we also note that it is now possible to compensate the extra harmonic trap
with a dipole trap in experiments \cite{Will},
which can further release the constraints. Finally if the filling deviates from unit filling per supercell,
holes or doublons can arise in the superlattice, which can scatter and couple to the collective excitations. New phenomena
can be generated due to such scattering and coupling, and we refer the reader for possible new phenomena to future investigations.

\section*{Acknowledgments}
This work is dedicated to Prof. Lorenz Cederbaum on the occasion of his 70th birthday. The authors acknowledge the efforts
of Sven Schmidt and Xiangguo Yin in the initial stage of the work. L. Cao is also grateful to Antonio Negretti for inspiring
discussions on ion phonons and the conditions of realistic implementations. S.M and P.S gratefully acknowledge funding
by the Deutsche Forschungsgemeinschaft (DFG) in the framework of the
SFB 925 ''Light induced dynamics and control of correlated quantum
systems''.

\appendix

\section{ML-MCTDHX}

The Multi-Layer Multi-Configuration Time-Dependent Hartree method for
multicomponent quantum gases (ML-MCTDHX) \cite{MLX} constitutes a variational numerical
ab-initio method for investigating both the stationary properties and in particular
the non-equilibrium quantum dynamics of mixture ensembles
covering the weak and strong correlation regimes. Its multi-layer
feature enables us to deal with multispecies systems (e.g. Bose-Bose, Fermi-Fermi or Bose-Fermi mixtures),
multidimensional or mixed dimensional systems in an efficient
manner. The multiconfigurational expansion of the wavefunction
in the ML-MCTDHX method takes into account higher band effects which
renders this approach suitable for the investigation of systems
governed by temporally varying Hamiltonians, where the system can be
excited to higher bands especially during the dynamics. Finally within
the ML-MCTDHX approach the representation of the wavefunction is performed by variationally
optimal (time-dependent) single particle functions (SPFs) and expansion
coefficients ${A_{{i_1}...{i_S}}}(t)$ which makes the
truncation of the Hilbert space
optimal when employing the optimal time-dependent moving basis. The requirement for convergence
demands a sufficient number of SPFs such that the numerical
exactness of the method is guaranteed. Therefore, the number of SPFs has to be increased until the quantities
of interest acquire the corresponding numerical accuracy.

In a generic mixture system consisting of ${N_\sigma }$ atoms (bosons or fermions) of
species $\sigma  = 1,2,...,S$ the main concept of the ML-MCTDHX
method is to solve the time-dependent Schr\"{o}dinger equation
$i {|\dot{\Psi}}\rangle    = \widehat {\rm H}\left| \Psi  \right\rangle$
as an initial value problem, $|{\Psi (0)}\rangle  = \left| {{\Psi _0}} \right\rangle$, by expanding the total wave-function in
terms of Hartree products
\begin{equation}
\begin{split}
\label{eq:3}\left| {\Psi (t)} \right\rangle  = \sum\limits_{{i_1} =
1}^{{M_1}} {\sum\limits_{{i_2} = 1}^{{M_2}} {...\sum\limits_{{i_S} =
1}^{{M_S}} {{A_{{i_1}...i{}_S}}(t)} } } \\\times\ket{{\psi
_{{i_1}}^{(1)}(t)}} ... \ket{{\psi _{{i_S}}^{(S)}(t)}}.
\end{split}
\end{equation}
Here each species state $\ket{\psi  _i^{(\sigma )}}$ ($i =
1,2,...,{M_\sigma }$) corresponds to a system of ${N_\sigma }$
indistinguishable atoms (bosons or fermions) and describes a many-body state of a subsystem composed
of $\sigma$-species particles. The expansion of each species state in terms of
bosonic or fermionic number states ${\ket{{\vec n (t)}}^\sigma }$ reads
\begin{equation}
\label{eq:10}\ket{\psi  _i^{(\sigma )}}  = \sum\limits_{\vec n
\parallel \sigma } {C_{i;\vec n }^\sigma (t)} {\left| {\vec n (t)}
\right\rangle ^\sigma },
\end{equation}
where each $\sigma $ atom can occupy ${m_\sigma }$ time-dependent
SPFs $\ket{\varphi _j^{(\sigma )}}$. The vector $\left| {\vec n } \right\rangle  = \left|
{{n_1},{n_2},...,{n_{{m_\sigma }}}} \right\rangle$ contains the
occupation number ${n_j}$ of the $j - th$ SPF that obeys the
constraint ${n_1} + {n_2} + ... + {n_{{m_\sigma
}}} = {N_\sigma }$. Note that for the bosonic case $n_j=0,1,2,...,N$ while for the fermionic case only $n_j=0,1$ are
permitted due to the Pauli exclusion principle.

In the present work, we focus on the case of a single fermionic species in one spatial dimension where
the ML-MCTDHX is equivalent to MCTDHF. To be self-contained, let us
briefly discuss the ansatz for the many-body wavefunction and the
procedure for the derivation of the equations of motion. The many-body wavefunction which
is a linear combination of time-dependent Slater determinants reads
\begin{equation}
\label{eq:10}\left| {\Psi (t)} \right\rangle  = \sum\limits_{\vec n
} {{C_{\vec n }}(t)\left| {{n_1},{n_2},...,{n_M};t} \right\rangle }.
\end{equation}
Here $M$ denotes the total number of SPFs and the
summation is performed over all possible combinations which retain the
total number of fermions. In the limit in which $M$
approaches the number of grid points the above expansion becomes numerically  
exact in the sense of a full configuration interaction approach. Another limiting case of 
the used expansion refers to the case that $M$ equals the number of particles, being referred to 
in the literature as Time-Dependent Hartree Fock (TDHF). The Slater determinants in (A3) can be expanded in terms of the creation 
operators $a_j^\dag (t)$ for the $j - th$ orbital ${\varphi _j}(t)$ as follows
\begin{equation}
\begin{split}
\label{eq:4}\left| {{n_1},{n_2},...,{n_M};t} \right\rangle  =
\frac{1}{{\sqrt {{n_1}!{n_2}!...{n_M}!} }}{\left( {a_1^\dag }
\right)^{{n_1}}}{\left( {a_2^\dag } \right)^{{n_2}}}\\\times...{\left(
{a_M^\dag } \right)^{{n_M}}}\left| {vac} \right\rangle,
\end{split}
\end{equation}
satisfying the standard fermionic anticommutation relations $\left[
{{a_i}(t),{a_j}^{\dag}(t)} \right]_- = {\delta _{ij}}$, etc.
To determine the time-dependent wave function $\left|
\Psi \right\rangle$, we have to find the equations of motion for the
coefficients ${{C_{\vec n }}(t)}$ and the orbitals (which are both
time-dependent). To derive the equations of motion for the mixture system one can employ various approaches
such as the Lagrangian, McLachlan or the
Dirac-Frenkel variational principle, each
of them leading to the same result. Following the Dirac-Frenkel
variational principle
\begin{equation}
\label{eq:5}{\bra{\delta \Psi}}{i{\partial _t} - \hat{ H}\ket{\Psi
}}=0,
\end{equation}
we can determine the time evolution of all the coefficients
${{C_{\vec n }}(t)}$ in the ansatz (A3) and the time dependence of
the orbitals $\left| {{\varphi _j}} \right\rangle $.
In this manner, we end up with a set
of $M$ non-linear integro-differential equations of motion for the
orbitals $\varphi_{j}(t)$, which are coupled to the $\frac{M!}{N!(M-N)!}$ linear
equations of motion for the coefficients $C_{\vec{n}}(t)$ . These equations are the
well-known MCTDHF equations of motion \cite{Alon,Alon1}.

Within our implementation, a discrete variable representation (DVR) scheme is applied, and in particular we adopt the sin-DVR, which
intrinsically implements hard-wall boundaries conditions. Furthermore, for the cases of three and five fermions, six and ten SPFs
have been used, respectively, i.e. the number of SPFs being twice the number of the particles. As it turned out, the number of major occupied natural orbitals
for both cases, reflecting the convergence of the simulation with respect to the number of SPFs, is equal to the number of
particles. This indicates that one just needs to use as many SPFs as there are fermions in order to reach a converged simulation. Finally,
in the simulation of the eleven-fermion case, only eleven SPFs have been used.

\section{Semiclassical equations}

We firstly re write the Hamiltonian of equation (1) in the Hubbard form. Upon the lowest-band Wannier states $\{|L\rangle_i,|R\rangle_i\}_{i=1}^N$,
we define a set of basis vectors of  $\{|S\rangle_i,|A\rangle_i\}_{i=1}^N$, where
$|S(A)\rangle_i\equiv(|L\rangle_i+(-)|R\rangle_i)/\sqrt{2}$ denote the corresponding symmetric (anti-symmetric) superposition within the $i$-th supercell.
By regarding $|S\rangle_i$/$|A\rangle_i$ as a pseudo-spin state of $|\downarrow\rangle$/$|\uparrow\rangle$, we introduce
the Pauli matrices $\sigma_\alpha$, with $\alpha=x,y,z$ for these basis vectors. 
We focus on the evolution of the system after the tilt is removed, and the corresponding Hamiltonian can then be expressed in this basis as 
\begin{equation}\label{hamhub}
\begin{split}
 \hat H = J\sum_{i=1}^N\sigma_z(i)+\sum_{i=1}^{N-1}V(S^{+}_iS^{-}_{i-1}+h.c.)\\+\sum_{i=1}^{N-1}V(S^{+}_iS^{+}_{i-1}+H.c.)
 +U(\sigma_x(1)-\sigma_x(N)),
 \end{split}
\end{equation}
where J refers to the intra-cell hopping strength, and V, U are determined by the interaction strength.
In this reduced Hamiltonian, we approximate the dipolar interaction by a nearest-neighbor interaction and neglect the inter-cell hopping,
which is valid within the weak interaction regime considered in this work \cite{Yin}.

Based on equation (B1), $|\Psi(t)\rangle$ can be solved analytically by perturbation theory. It turns out to be enough to use
the first order perturbation. In the perturbation treatment, we take the last two terms in equation (B.1) as a perturbation.
To zero-th order, the ground state is given by equation (2). The first-order correction of degenerate
perturbation theory gives that a set of low-lying excited states bunch into a band on top of the ground state,
and the eigenstates in this first excited band can be expressed as 
\begin{equation}\label{k-state}
 |k\rangle = \sum_{i=1}^{N}\sin(\frac{ki\pi}{N+1})|i\rangle_{-},
\end{equation}
where $|i\rangle_{-}=\big(\prod_{n\in [1,i-1]\bigcup[i+1,N]}|S\rangle_n\big)\times|A\rangle_i$. 
It can be shown that for the collective dynamics considered here, it is enough to focus on the ground state
and first excited band.
Without loss of generality, we can assume the wave function at time $\tau$, $i.e.$ when $V_{pt}$ vanishes, as
$|\Psi(\tau)\rangle=(\alpha|S\rangle_1+\beta|A\rangle_1)\times\prod_{i=2}^N|S\rangle_i$ with $|\alpha|^2+|\beta|^2=1$. 
Then $|\Psi(t-\tau)\rangle$ becomes 
\begin{equation}\label{wfunc}
\begin{split}
 |\Psi(t-\tau)\rangle=&e^{-i\epsilon_0 (t-\tau)}|G\rangle\langle G|\Psi(\tau)\rangle
 \\&+\sum_{k=1}^N e^{-i\epsilon_k (t-\tau)}|k\rangle\langle k|\Psi(\tau)\rangle
 \\=&\alpha e^{-i\epsilon_0 (t-\tau)}|G\rangle\\
 &+\sum_{k=1}^N\beta sin(\frac{k\pi}{N+1})e^{-i\epsilon_k (t-\tau)}|k\rangle.
\end{split}
\end{equation}

In the basis of $\{|S\rangle_i,|A\rangle_i\}_{i=1}^N$, $\delta\hat\rho_i$ becomes $\sigma_x(i)$. Substituting this expression
and equation (B1) into equation (4), we obtain
\begin{equation}\label{commut}
\begin{split}
 [[\delta\hat\rho_i,\hat H],\hat H]=4J^2\sigma_x(i)-4JV\sigma_z(i)\\\times(\sigma_x(i+1)+\sigma_x(i-1)).
 \end{split}
\end{equation}
Further using equation (B3) for the average $<...>$,
it can be proven that $\langle\sigma_z(i)\sigma_x(i\pm 1)\rangle=\langle\sigma_x(i\pm 1)\rangle$. It is then
straightforward to obtain the time derivative equation (5).

{}

\end{document}